\documentclass[12pt]{article}    
\usepackage[utf8]{inputenc}
\usepackage{hyperref}
\usepackage{natbib}
\usepackage{graphics}
\usepackage{epsfig}
\usepackage{amsmath}
\usepackage{amsfonts}
\usepackage{amssymb}
\usepackage{epstopdf}
\usepackage{placeins}
\usepackage{float}
\usepackage{tabularx}
\usepackage{array}
\usepackage{xcolor}

\usepackage{booktabs}
\usepackage{multirow}
\usepackage{graphicx}

\begin{document}
\newtheorem{definition}{Definition}
\newtheorem{theorem}{Theorem}
\newtheorem{example}{Example}
\newtheorem{corollary}{Corollary}
\newtheorem{lemma}{Lemma}
\newtheorem{proposition}{Proposition}
\newtheorem{remark}{Remark}
\newenvironment{proof}{{\bf Proof:\ \ }}{\qed}
\newcommand{\qed}{\rule{0.5em}{1.5ex}}
\newcommand{\bfg}[1]{\mbox{\boldmath $#1$\unboldmath}}

\numberwithin{proposition}{section}
\numberwithin{equation}{section} \numberwithin{theorem}{section}
\numberwithin{example}{section} \numberwithin{definition}{section}
\numberwithin{lemma}{section} \numberwithin{remark}{section}

\title{Measuring AI harms with multidimensional Lorenz Zonoids}

\author{%
\parbox{0.95\textwidth}{%
\centering
Paolo Giudici\textsuperscript{a},
Jos\'e Mar\'{\i}a Sarabia\textsuperscript{b},
Sofia Vei\textsuperscript{c}
\\[0.7em]
\normalsize
\textsuperscript{a}Department of Economics and Management,
University of Pavia, Italy
\\[0.25em]
\textsuperscript{b}Department of Economics and SANFI Institute,
University of Cantabria, Santander, Spain
\\[0.25em]
\textsuperscript{c}School of Informatics,
Aristotle University of Thessaloniki,
54124 Thessaloniki, Greece
}}

\date{\today}

\maketitle

\begin{abstract}
While AI systems increasingly shape high-stakes societal domains, their governance is limited by the lack of risk management methods that operate on real harms, taking their severity, and not only their likelihood, into account. As  a consequence, AI risk management  models remain compliance-driven and provider-centric, offering limited insight into how harms are dangerous, and on what should be the priority of intervention. The problem is amplified by the nature of harm data which are typically ordinal and multidimensional. 
To solve the problem, and offer an effective risk assessment methodology, in this paper we propose to model harm data by means of Lorenz Zonoids and  Gini indices. To this aim we propose to extend them in a multidimensional setting,  and show how to practically calculate them for a real AI incident data repository, provided by the Massachusetts Institute of Technology. The empirical findings indicate that environmental, infrastructure,
property, physical, and democracy-related harms attain the highest
values under the two multidimensional Gini indices and therefore
exhibit the strongest concentration in their joint direct, indirect,
and inferred severity-frequency distributions. These concentration
patterns may help identify categories that warrant closer examination
when mitigation priorities are determined.
\end{abstract}


\section{Introduction}

The growing adoption of Artificial Intelligence (AI) in high-stakes domains has intensified the need for evaluation frameworks capable of assessing dimensions that go beyond predictive accuracy. Within this context, the SAFE (Sustainability, Accuracy, Fairness, Explainability) framework has emerged as an integrated approach for measuring and managing AI trustworthiness, particularly in economic and financial applications. The framework was initially proposed by \cite{Giudici2023safe}: the authors introduced a set of statistical indicators designed to evaluate whether AI systems satisfy regulatory and operational requirements for trustworthy deployment. In particular, the framework translates principles embedded in regulations such as the European Union AI Act into measurable dimensions that can be monitored over time. As a related contribution, \cite{Giudici2024artificial} introduce Key AI Risk Indicators (KAIRI) to quantify and monitor AI-related risks in finance. In this setting, SAFE metrics are interpreted as governance tools capable of supporting model auditing, regulatory compliance, and operational risk control. These developments position SAFE AI within the broader literature on trustworthy and responsible AI \citep{Liu2022trustworthy, Kaur2022trustworthy, Cheng2021socially}, which emphasizes robustness, transparency, accountability, and fairness as central requirements for AI deployment in critical environments.

 In this work, we distinguish \textit{AI harms}, realized negative impacts (e.g., physical, psychological, social, or economic loss), from \textit{AI risks}, which describe uncertainty about whether and how such harms may occur. While the latter can be assessed using AI evaluation tools, such as those in the mentioned SAFE and KAIRI frameworks, the former requires the analysis of actual harm data. A comprehensive AI risk management framework needs both the likelihood and the severity of AI harms: while the former can be assessed by AI evaluation tools, using data internal to the AI provider, the latter requires the analysis of external impacts.

Indeed, despite rapid regulatory and governance developments, most AI risk assessment frameworks still adopt an internal, compliance-oriented perspective, emphasizing predefined checklists and provider-centric controls \cite{hoffman2023outside}. This can under-represent affected stakeholders and miss emergent harms \cite{lancaster2024s}. Coarse severity schemes (e.g., binary) also blur the distinction between minor and high-stakes failures, a key issue for risk-based regulation such as the European AI Act \cite{smuha2025european, jariwala2024comparative, novelli2024taking}.

A key obstacle is that severity information in incident repositories is often noisy or unavailable. When numerical scores exist, they may be inconsistently assigned across sources, and annotators may disagree due to incomplete information and epistemic uncertainty. Many repositories therefore report ordinal severity (e.g., minor, moderate, severe), which supports ranking without imposing unjustified distance assumptions \cite{robinson2024likert}. Similar ordinal approaches are used in cyber and operational risk where quantitative evidence is limited \cite{giudici2021cyber}.

Motivated by these gaps, we pose the following research question: How can harms be prioritized and, thus, mitigated,  when precise numerical severity data are unavailable or unreliable? 

To answer this question, we need an ordinal measure of severity that is also multidimensional.

\cite{VEI2026104587} proposed a unidimensional measure of severity, named AIH, that can be employed when harms are measured in one dimension. In reality, however, harms are often measured across more than one severity dimension. For example, in the MIT AI Incident Tracker, each incident is not described by one severity value only. Instead, each incident receives severity ratings for ten harm dimensions, and each of these dimensions is further evaluated across three pathways: direct, indirect, and inferred. Thus, the same incident may simultaneously carry different severity information depending on whether one observes the immediate harm, a secondary harm, or a harm inferred from the incident description.

This structure cannot be fully represented by a single unidimensional AIH score without losing information. A pathway-wise AIH analysis remains useful as a preliminary diagnostic, because it shows how each harm dimension behaves when direct, indirect, and inferred severities are considered separately. Yet the substantive MIT problem is multidimensional: the three pathways form a vector of ordinal severity information for each harm dimension. The  aim of the present paper is therefore to move from the unidimensional AIH logic to a multidimensional representation that can preserve the joint structure of these severity pathways.

To measure harm severity in the multidimensional case, we propose multidimensional lorenz zonoids, the extension of the Gini index to multiple dimensions, which was introduced  theoretically, but never made explicit and computable.

We empirically demonstrate the framework on the MIT dataset, showing
that environmental, infrastructure, property, physical, and
democracy-related harms attain the highest values under the two
multidimensional Gini indices. Accordingly, these categories exhibit
the strongest concentration in their joint Direct--Indirect--Inferred
severity-frequency distributions and may warrant closer examination
when mitigation strategies are considered.

We finally remark that the main contribution of this paper is to show how to calculate a Lorenz Zonoid for a multidimensional response variable, to provide an AI  harm assessment metric. When a multidimensional response variable is considered, \cite{Auricchio2026rank} suggested using a whitening projection, which converts it into a univariate functional, for which a unidimensional Lorenz Zonoid can be calculated. In this paper we would like to  avoid the use of projections, for greater transparency. To this aim, we will show how to obtain a  multidimensional Lorenz Zonoid for a multidimensional response variable.

\section{Extending the Lorenz curve to higher dimensions}

A key aspect of the methodology in this paper is the use of multivariate data concentration measures. These measures involve extending the classical Lorenz curve (see \cite{arnold2018majorization}) to higher dimensions. This kind of extension has been an active research area whose solution has not been straightforward, partly due to a lack of adequate analytical tools. The absence of a convincing definition of the quantile function in dimensions greater than or equal to two has been one of the main obstacles. We will call the version of the Lorenz curve in dimensions greater than or equal to two the Multivariate Lorenz Surface (MLS). The first attempts to extend the Lorenz curve to dimensions equal to or greater than two were due to \cite{taguchi1972two} and \cite{lunetta1972di}.

Subsequently, in the two-dimensional case, \cite{arnold1983pareto} proposed an alternative definition of a Lorenz surface for bivariate distributions with marginal distributions $F_1$ and $F_2$, indexed by $(x_1,x_2)$ and with joint pdf $f_{12}(x_1,x_2)$. The points of this surface are given by
\begin{equation}\label{arnold1983}
\left(F_1(x_1),F_2(x_2),\frac{1}{E(X_1X_2)}\int_0^{x_1}\int_0^{x_2}u_1u_2f_{12}(u_1,u_2)\,du_1\,du_2\right).
\end{equation}
The extension of (\ref{arnold1983}) to the case $p>2$ is straightforward.

The following methodology establishes the concept of the Lorenz zonoid, proposed by \cite{koshevoy1995multivariate}, \cite{koshevoy1996lorenz}, and \cite{koshevoy1997zonoid}. In \cite{koshevoy1997zonoid}, this concept was introduced as a geometric method that allows the idea of the Lorenz curve to be extended to dimensions higher than one, that is, $p\ge 1$. Denote by ${\cal L}_+^p$ the set of all $p$-dimensional non-negative random vectors $\mathbf{X}$ with $E(\mathbf{X})<\infty$, and let $\Psi^{(p)}$ denote the class of all measurable mappings from $\mathbb{R}^p_+$ to $[0,1]$. The Lorenz zonoid $LZ(\mathbf{X})$ of the random vector $\mathbf{X}\in {\cal L}_+^p$ with distribution $F(\mathbf{x})$ is defined as the set of points
\begin{eqnarray*}
LZ(\mathbf{X})&=&\left\{\left(\int\psi(\mathbf{x})\,dF_{\mathbf{X}}(\mathbf{x}),\frac{\int x_1\psi(\mathbf{x})\,dF_{\mathbf{X}}(\mathbf{x})}{E(X_1)},
\dots,\frac{\int x_p\psi(\mathbf{x})\,dF_{\mathbf{X}}(\mathbf{x})}{E(X_p)}\right)\right\}\nonumber\\
&=&\left\{\left(E(\psi(\mathbf{X})),\frac{E(X_1\psi(\mathbf{X}))}{E(X_1)},\dots,\frac{E(X_p\psi(\mathbf{X}))}{E(X_p)}\right):\psi\in\Psi^{(p)}\right\}.
\end{eqnarray*}
In the case $p=1$, the zonoid corresponds to the points in the plane that lie between the Lorenz curve $L(x)$ and the reverse Lorenz curve $1-L(1-x)$. The empirical version of the zonoid will be studied in the next section.

More recently, two new types of MLS have been proposed, based on different methodologies: on the one hand, copula-based methodologies, and, on the other, methodologies based on modern optimal transportation theory (OTT).

The copula-based methodology was proposed by \cite{grothe2022multivariate}. In this approach, the inverse functions of the marginal Lorenz curves are used, and the joint distribution is obtained by means of a copula, either an empirical one or a copula from a parametric family.

The OTT-based methodology has given rise to two proposals: \cite{fan2024multidimensional} and \cite{hallin2025multiple}. In both cases, OTT is used to obtain a suitable definition of the multivariate quantile function. In this context, \cite{fan2024multidimensional} defines a quantile function satisfying four basic properties: $Q_\mathbf{X}$ is invertible, $Q_\mathbf{X}^{-1}(\mathbf{X})$ is uniformly distributed on $[0,1]^p$, the map $Q(\mathbf{u})$ is cyclically monotone, and, consequently, its gradient is a convex potential. Let $\mathbf{U}$ be uniformly distributed on $[0,1]^p$, let $\mathbf{X}$ be a random vector on $\mathbb{R}_+^p$, let $\tilde{\mathbf{X}}=(X_1/\mu_1,\dots,X_p/\mu_p)$, and let $Q_{\tilde{\mathbf{X}}}$ be the vector quantile of $\tilde{\mathbf{X}}$. Then, the so-called Lorenz allocation map of $\mathbf{X}$ is the vector-valued function
$$
L_{\mathbf{X}}(\mathbf{u})=\int_{0}^{u_1}\cdots \int_{0}^{u_p}Q_{\tilde{\mathbf{X}}}(\mathbf{v})\,d\mathbf{v}.
$$

The second OTT-based proposal, by \cite{hallin2025multiple}, makes use of the notion of the center-outward quantile function and the center-outward quantile region. The center-outward quantile region containing the proportion $\tau$ of the most central values of $\mathbf{X}$ is defined, for $\tau\in(0,1)$, as
$$
\mathbb{C}_{\mathbf{X}\pm}(\tau)=\{\mathbf{x}\in \mathbb{R}^p:\;||\mathbf{F}_{\mathbf{X}\pm}(\mathbf{x})||\le \tau\}=\mathbf{Q}_{\mathbf{X}\pm}(\tau \overline{\mathbb{S}}_p),
$$
where $\overline{\mathbb{S}}_p$ is the closed unit ball. Its inverse map is the so-called center-outward quantile function, given by $\mathbf{Q}_{\mathbf{X}\pm}$. This function has the nature of a potential, extending the notion of the primitive of the univariate cdf $F$. In this way, the absolute center-outward Lorenz function of $\mathbf{X}$ is the mapping
$$
L_{\mathbf{X}\pm}(\tau):=E\left[\mathbf{X}\,\mathbf{1}_{\{\mathbf{X}\in \mathbb{C}_{\mathbf{X}\pm}(\tau)\}}\right],\qquad 0\le \tau\le 1.
$$

\section{Methodological Issues}
In this section, we present the methodological contribution of the paper. After reviewing the different proposals for Lorenz surfaces in the previous section, we first present the empirical version of the Lorenz zonoid. The most noteworthy aspect of this part is a new way to empirically obtain the Lorenz zonoid, which had not been previously explored. Next, we present three multivariate concentration indices that will be used in the empirical part of the work.

\subsection{Empirical Lorenz zonoids}
\begin{figure}[H]
\begin{center}
\includegraphics[scale=1.2]{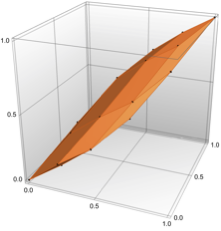}
\caption{The Lorenz zonoid for the bivariate dataset (\ref{dataset1})}\label{graphic1}
\end{center}
\end{figure}

\begin{figure}[H]
\begin{center}
\includegraphics[scale=0.8]{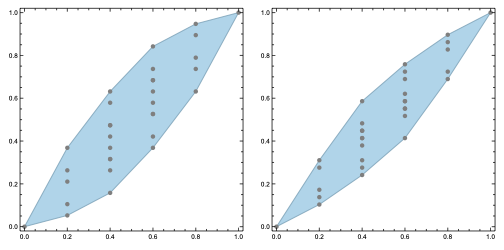}
\caption{The Lorenz zonoid for the marginal univariate dataset (\ref{dataset1})}\label{graphic2}
\end{center}
\end{figure}

In this section we propose the empirical version of the zonoid presented in the previous section. According to \cite{koshevoy1996lorenz}, the Lorenz zonoid of an empirical distribution $F_A$, with $A\subset \mathbb{R}^{n\times p}$, is defined as
$$
LZ(F_A)=\left\{
\mathbf{x}\in \mathbb{R}^{p+1},\;\mathbf{x}=\sum_{i=1}^{n}\alpha_i\mathbf{a}_i,\;\;0\le \alpha_i\le 1,\;\forall i
\right\},
$$
where
$$
\mathbf{a}_i=\left(\frac{1}{n},\frac{x_{i1}}{\sum_{l=1}^nx_{l1}},\dots,\frac{x_{ip}}{\sum_{l=1}^nx_{lp}}\right).
$$
Then, $LZ(F_A)$ is the convex hull of the points
$$
\sum_{i=1}^{n}\delta_i\mathbf{a}_i,\;\;\delta_i\in\{0,1\},\;\;i=1,2,\dots,n.
$$
To calculate the empirical Lorenz Zonoid, we now consider the construction suggested in 
\cite{arnold2011inequality}; see also \cite{marshall1979inequalities} and \cite{arnold2018majorization}. More formally,  we propose to construct the empirical Lorenz zonoid as a convex set in $\mathbb{R}^{p+1}$ that considers all possible combinations of population shares and the corresponding totals. 
To compute the zonoid, we consider all subsets $G_k$ of the
$n$ observations having size $k$, for $k=0,1,\dots,n$, where
$G_0=\varnothing$. For each $k$, the number of such subsets is
${n\choose k}$. Therefore, the total number of candidate subset-sum
points is
\[
\sum_{k=0}^{n}{n\choose k}=2^n,
\]
independently of the number of dimensions $p$. The empty subset
$G_0$ produces the origin $(0,\dots,0)$, which is part of the
Lorenz zonoid but is omitted from Table~\ref{Table1} because it is
trivial. For a subset $G_k$ of size $k$, the quantity $k/n$ is the
corresponding population share.
For a given subset, we consider the vector in $\mathbb{R}^{p+1}$
\begin{equation}\label{NFormula}
\left(\frac{k}{n},\frac{\sum_{i\in G_k}x_{i1}}{\sum_{i=1}^{n}x_{i1}},\dots,\frac{\sum_{i\in G_k}x_{ip}}{\sum_{i=1}^nx_{ip}}\right).
\end{equation}
To illustrate our proposal, we now present a numerical example of how to compute the zonoid in the case $p=2$ using the methodology described above. We consider the dataset
\begin{equation}\label{dataset1}
\mathbf{x}=\{(1,4),(2,3),(4,5),(5,8),(7,9)\}.
\end{equation}
Table \ref{Table1} includes the elements necessary to compute the zonoid points. Since $n=5$, we have groups of sizes 1 to 5. The second column lists the different groups, and the third column gives the zonoid points according to formula (\ref{NFormula}). Figure \ref{graphic1} shows the zonoid, obtained as the convex hull of these points. Figure \ref{graphic2}, on the other hand, shows the marginal zonoids corresponding to the first and second components of the data.

\begin{table}[H]
    \centering
    \footnotesize
    \renewcommand{\arraystretch}{1.2}
    \setlength{\tabcolsep}{4pt}
\begin{tabular}{|c|c|c|}
\hline
 \bf{Size} & \bf{Groups} & \bf{Point of the Lorenz zonoid} \\
\hline
  $1$   & $\{(1,4)\}$ & $(\frac{1}{5},\frac{1}{19},\frac{4}{29})$ \\
        & $\{(2,3)\}$ & $(\frac{1}{5},\frac{2}{19},\frac{3}{29})$ \\
        & $\{(4,5)\}$ & $(\frac{1}{5},\frac{4}{19},\frac{5}{29})$ \\
        & $\{(5,8)\}$ & $(\frac{1}{5},\frac{5}{19},\frac{8}{29})$ \\
        & $\{(7,9)\}$ & $(\frac{1}{5},\frac{7}{19},\frac{9}{29})$ \\
\hline
 $2$    & $\{(1,4),(2,3)\}$ & $(\frac{2}{5},\frac{3}{19},\frac{7}{29})$ \\
        & $\{(1,4),(4,5)\}$ & $(\frac{2}{5},\frac{5}{19},\frac{9}{29})$ \\
        & $\{(1,4),(5,8)\}$ & $(\frac{2}{5},\frac{6}{19},\frac{12}{29})$ \\
        & $\{(1,4),(7,9)\}$ & $(\frac{2}{5},\frac{8}{19},\frac{13}{29})$ \\
        & $\{(2,3),(4,5)\}$ & $(\frac{2}{5},\frac{6}{19},\frac{8}{29})$ \\
        & $\{(2,3),(5,8)\}$ & $(\frac{2}{5},\frac{7}{19},\frac{11}{29})$ \\
        & $\{(2,3),(7,9)\}$ & $(\frac{2}{5},\frac{9}{19},\frac{12}{29})$ \\
        & $\{(4,5),(5,8)\}$ & $(\frac{2}{5},\frac{9}{19},\frac{13}{29})$ \\
        & $\{(4,5),(7,9)\}$ & $(\frac{2}{5},\frac{11}{19},\frac{14}{29})$ \\
        & $\{(5,8),(7,9)\}$ & $(\frac{2}{5},\frac{12}{19},\frac{17}{29})$ \\
\hline
  $3$   & $\{(1,4),(2,3),(4,5)\}$ & $(\frac{3}{5},\frac{7}{19},\frac{12}{29})$ \\
        & $\{(1,4),(2,3),(5,8)\}$ & $(\frac{3}{5},\frac{8}{19},\frac{15}{29})$ \\
        & $\{(1,4),(2,3),(7,9)\}$ & $(\frac{3}{5},\frac{10}{19},\frac{16}{29})$ \\
        & $\{(1,4),(4,5),(5,8)\}$ & $(\frac{3}{5},\frac{10}{19},\frac{17}{29})$ \\
        & $\{(1,4),(4,5),(7,9)\}$ & $(\frac{3}{5},\frac{12}{19},\frac{18}{29})$ \\
        & $\{(1,4),(5,8),(7,9)\}$ & $(\frac{3}{5},\frac{13}{19},\frac{21}{29})$ \\
        & $\{(2,3),(4,5),(5,8)\}$ & $(\frac{3}{5},\frac{11}{19},\frac{16}{29})$ \\
        & $\{(2,3),(4,5),(7,9)\}$ & $(\frac{3}{5},\frac{13}{19},\frac{17}{29})$ \\
        & $\{(2,3),(5,8),(7,9)\}$ & $(\frac{3}{5},\frac{14}{19},\frac{20}{29})$ \\
        & $\{(4,5),(5,8),(7,9)\}$ & $(\frac{3}{5},\frac{16}{19},\frac{22}{29})$ \\
\hline
  $4$   & $\{(1,4),(2,3),(4,5),(5,8)\}$ & $(\frac{4}{5},\frac{12}{19},\frac{20}{29})$ \\
        & $\{(1,4),(2,3),(4,5),(7,9)\}$ & $(\frac{4}{5},\frac{14}{19},\frac{21}{29})$ \\
        & $\{(1,4),(2,3),(5,8),(7,9)\}$ & $(\frac{4}{5},\frac{15}{19},\frac{24}{29})$ \\
        & $\{(1,4),(4,5),(5,8),(7,9)\}$ & $(\frac{4}{5},\frac{17}{19},\frac{26}{29})$ \\
        & $\{(2,3),(4,5),(5,8),(7,9)\}$ & $(\frac{4}{5},\frac{18}{19},\frac{25}{29})$ \\
\hline
 $5$    & $\{(1,4),(2,3),(4,5),(5,8),(7,9)\}$  & $(1,1,1)$ \\
\hline
\end{tabular}
\caption{Computation of the zonoid for $p=2$ and $n=5$ for the data set (\ref{dataset1}) }\label{Table1}
\end{table}

\subsection{Multivariate Gini indices}

In this section we introduce some multivariate Gini indices that will be used in the application. Let $A=\{a_{is}\}$ be an $n\times d$ data matrix ($i=1,2,\dots,n$ and $s=1,2,\dots,d$), where $d$ is the dimension and $n$ is the number of data, and $a_i$ is its $i$th row. We denote as $F_A$ the $d$ variate empirical distribution that puts equal mass $1/n$ to each $a_i$. We define the distance-Gini mean difference as,
\begin{equation}\label{gini1}
M_D(F_A)=\frac{1}{2n^2d}\sum_{i=1}^{n}\sum_{j=1}^{n}\left(\sum_{s=1}^{d}(a_{is}-a_{js})^2\right)^{1/2}.
\end{equation}
The classical univariate Gini mean difference has the following multivariate extension. Then, the distance-Gini mean difference is,
\begin{equation}\label{gini2}
M_D(F)=\frac{1}{2d}\int_{\mathbb{R}^d}\int_{\mathbb{R}^d} \|x-y\|dF(x)dF(y),
\end{equation}
where $\|\cdot\|$ denotes the Euclidean distance in $\mathbb{R}^d$. In the case of an empirical distribution, $M_D(F_A)$ is given by (\ref{gini1}) and $R_D(F_A)$ by,
\begin{equation}\label{gini3}
R_D(F_A)=\frac{1}{2dn^2}\sum_{j=1}^{n}\sum_{i=1}^{n}\left(\sum_{s=1}^{d}\frac{(a_{is}-a_{js})^2}{\bar{a}_s^2}\right)^{1/2}
\end{equation}
Properties of (\ref{gini2}) and (\ref{gini3}) are discussed in \cite{koshevoy1997multivariate}.

Another important index is based on the expanded volume of the lift zonoid and is named the volume-Gini index difference. It is defined as,
\begin{equation}\label{volumeGini}
M_V(F_A)=\frac{1}{2^d-1}\sum_{s=1}^{d}\frac{1}{n^{s+1}}\sum_{1\le i_1<\dots <i_{s+1}\le n}\;\sum_{1\le r_1<\dots<r_s\le d}|\mbox{det}(\mathbf{1},A_{i_1,\dots,i_{s+1}}^{r_1,\dots,r_s})|
\end{equation}
where $\mathbf{1}$ is a column of ones and $A_{i_1,\dots,i_{s+1}}^{r_1,\dots,r_s}$ is the matrix obtained from the rows $i_1,\dots,i_{s+1}$ and the columns $r_1,\dots,r_s$ of the data matrix.

In order to enable a comparison between indices (\ref{gini3}) and (\ref{volumeGini}), we establish upper bounds for both quantities. Let ${\cal F}^d$ be the class of probability distributions on $\mathbb{R}^d$ with finite mean vectors. For $F\in{\cal F}^d$, according to Theorem 4.1 in \cite{koshevoy1997multivariate}, we have
\begin{equation}\label{bound1}
M_D(F)<\frac{1}{d}\sum_{j=1}^d\mu_j(F),\;\;R_D(F)<1,
\end{equation}
hold and the bounds are sharp. On the other hand, according to Proposition 5.2 in \cite{koshevoy1997multivariate}, if $F\in{\cal F}^d$, we have
\begin{equation}\label{bound2}
M_V(F)
<
\frac{1}{2^d-1}
\sum_{\varnothing\neq K\subset\{1,\dots,d\}}
\prod_{i\in K}\mu_i
\le
\frac{1}{2^d-1}
\left(
(\max_i \mu_i+1)^d-1
\right),
\end{equation}
and the first inequality cannot be improved.

Therefore, according to (\ref{bound1}), index (\ref{gini3}) can be used directly and (\ref{volumeGini}) must be corrected by the last bound included in (\ref{bound2}).

\section{MIT Incident Data}\label{sec:mit}

This section describes the MIT AI Incident Tracker data used in the empirical analysis. We use the export dated 10 June 2026, which contains 1498 incidents involving AI, algorithmic, or automated systems. Each row corresponds to one incident and includes descriptive fields, such as the incident title, summary, description, source URL, alleged deployer, alleged developer, alleged harmed parties, year, and broader classification fields. The export also assigns each incident to a domain and subdomain. In this version of the data, the largest domains are malicious actors (567 incidents), AI system safety, failures, and limitations (314 incidents), discrimination and toxicity (243 incidents), misinformation (196 incidents), privacy and security (112 incidents), human-computer interaction (42 incidents), and socioeconomic and environmental harms (24 incidents). Additional columns record confidence, missing-information notes, number-of-people fields, behaviour type, EU AI Act risk classification, and national-security annotations. These fields are useful for describing and interpreting the dataset, but they are not the variables used to compute the ordinal AIH scores reported below.

The relevant analytical variables are the incident-level severity annotations attached to ten harm dimensions: physical, infrastructure, property, financial, environmental, malicious content, differential treatment, civil rights, democracy, and privacy. For each harm dimension, the export reports three severity pathways, labelled direct, indirect, and inferred. Direct severity refers to harms recorded as directly associated with the incident; indirect severity records secondary or mediated harms; inferred severity records additional severity assessments inferred from the incident information. This gives 30 severity columns per incident, or 44940 incident--harm--pathway severity observations before aggregation.
Table~\ref{tab:mit_raw_example} gives an illustrative extract of the raw MIT export before aggregation. Each row is an incident, while the severity columns already encode harm-specific ratings by pathway. The full export contains many additional descriptive and justification fields; the table shows only a small subset of columns to make the raw structure visible.

\begin{table}[H]
\centering
\scriptsize
\renewcommand{\arraystretch}{1.15}
\setlength{\tabcolsep}{3pt}
\begin{tabularx}{\textwidth}{@{}p{0.08\textwidth}p{0.10\textwidth}X p{0.17\textwidth}>{\centering\arraybackslash}p{0.10\textwidth}>{\centering\arraybackslash}p{0.12\textwidth}>{\centering\arraybackslash}p{0.11\textwidth}@{}}
\toprule
\textbf{ID} & \textbf{Year} & \textbf{Domain} & \textbf{Harmed parties} & \textbf{Physical direct} & \textbf{Malicious direct} & \textbf{Privacy inferred} \\
\midrule
1 & 2015 & Discrimination \& Toxicity & children & 1 & 3 & 1 \\
2 & 2018 & AI system safety, failures, \& limitations & warehouse-workers & 2 & 1 & 1 \\
3 & 2018 & AI system safety, failures, \& limitations & airplane-passengers; airplane-crew & 4 & 1 & 1 \\
\bottomrule
\end{tabularx}
\caption{Illustrative raw MIT rows before preprocessing. The original dataset is incident-level: descriptive fields, harmed-party fields, domain labels, and pathway-specific severity ratings appear in the same row. Only selected columns are shown here for readability.}
\label{tab:mit_raw_example}
\end{table}

The MIT severity values are treated as ordinal ratings. In the export used here, the observed rating scale runs from 0 to 5. These labels define an ordering from lower to higher severity, but they do not by themselves justify treating the distance between adjacent levels as equal. In the pathway-specific AIH pre-analysis, the full observed ordered scale, 0--5, is used. In the zonoid and concentration-index analysis, the value 0 is treated as the absence of positive recorded severity for the corresponding harm--pathway combination. The five positive severity levels, 1--5, are then used as ordered groups, and the zonoids are computed from the corresponding pathway-specific frequency counts or frequency shares, not from the severity labels as cardinal scores.

\section{Empirical results}

This section reports the empirical analysis in two steps. First, we use AIH as a pathway-specific pre-analysis to examine the ordinal severity distribution separately for the Direct, Indirect, and Inferred pathways. Second, we use empirical Lorenz zonoids and the associated multivariate concentration indices to analyse the joint severity-frequency structure across the three pathways.

\subsection{Univariate analysis: pathway-specific AIH}

Before constructing the multidimensional Lorenz zonoids, we first examine how severity is distributed within each pathway separately. This preliminary step uses the AIH measure \cite{VEI2026104587}, which is designed for ordered severity information and avoids imposing artificial numerical distances on ordinal categories. The harmed-party field is retained for descriptive context, but it is not used in the AIH calculation because the MIT export does not attach severity ratings to harmed-party groups. Instead, severity ratings are already attached to harm dimensions at the incident level. We therefore aggregate the data into triples of the form
\[
(\text{harm category},\text{severity rating},\text{frequency}).
\]
For each pathway and harm category, we count how many incidents fall into each ordered severity rating. Let $F_k$ denote the cumulative share of incidents up to rating level $k$, and let $k/m$ denote the normalized rank of that rating among the $m$ ordered severity levels. The AIH score is then obtained as the area under the corresponding ordered Lorenz-type curve. Higher AIH values indicate a stronger concentration of a harm category toward the upper part of the ordered severity scale.

Table~\ref{tab:mit_pathway_aih} compares AIH values across the three pathways. In the direct pathway, the largest value is obtained for malicious content harms ($AIH=0.3588$), followed by civil rights ($AIH=0.3162$), privacy ($AIH=0.3112$), financial harms ($AIH=0.3063$), and differential treatment ($AIH=0.3047$). This means that, within direct harms, malicious content has the strongest ordinal concentration toward higher severity ratings. At the other end, environmental harms ($AIH=0.2499$) and infrastructure harms ($AIH=0.2514$) are close to the lower part of the observed range, indicating that their direct severity ratings are more heavily concentrated at low levels. The ranking is not identical across pathways: malicious content is also highest for indirect harms ($AIH=0.2981$), whereas financial harm is highest for inferred harms ($AIH=0.3678$). Civil rights and privacy remain among the higher-AIH categories in the inferred pathway. This pattern suggests that the direct, indirect, and inferred annotations capture related but distinct severity structures, which is why the later zonoid step should combine the pathways rather than collapse them prematurely.

\begin{table}[htbp]
\centering
\footnotesize
\renewcommand{\arraystretch}{1.15}

\begin{tabular}{lccc}
\toprule
\textbf{MIT harm category} & \textbf{Direct AIH} & \textbf{Indirect AIH} & \textbf{Inferred AIH} \\
\midrule
Malicious Content & 0.3588 & 0.2981 & 0.3588 \\
Civil Rights & 0.3162 & 0.2971 & 0.3583 \\
Privacy & 0.3112 & 0.2850 & 0.3505 \\
Financial & 0.3063 & 0.2760 & 0.3678 \\
Differential Treatment & 0.3047 & 0.2897 & 0.3253 \\
Democracy & 0.2740 & 0.2758 & 0.2998 \\
Physical & 0.2730 & 0.2609 & 0.2769 \\
Property & 0.2595 & 0.2536 & 0.2625 \\
Infrastructure & 0.2514 & 0.2504 & 0.2524 \\
Environmental & 0.2499 & 0.2499 & 0.2501 \\
\bottomrule
\end{tabular}

\caption{MIT ordinal AIH values by harm category and severity pathway.}
\label{tab:mit_pathway_aih}

\end{table}

\begin{figure}[H]
\centering
\includegraphics[width=1\textwidth]{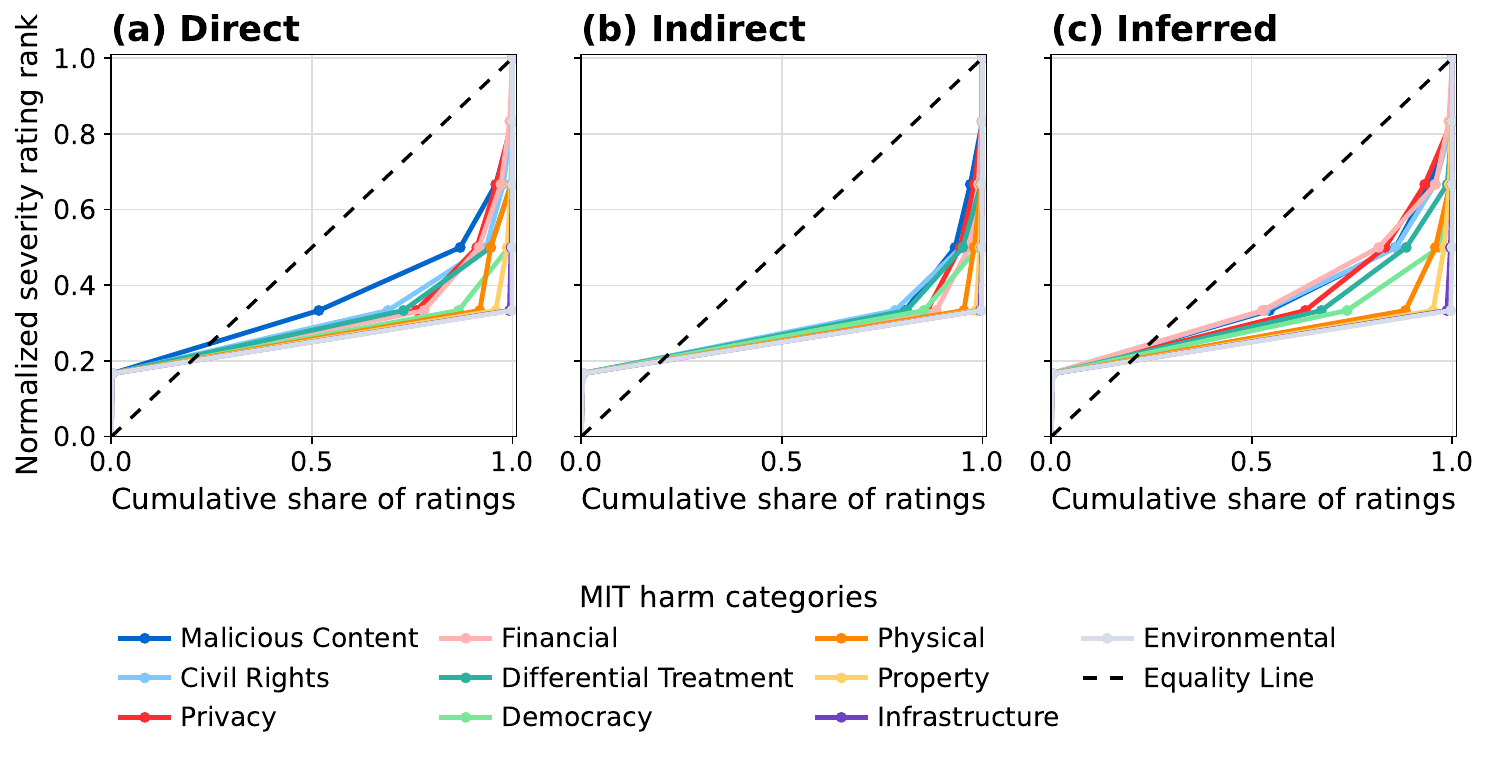}
\caption{Ordered Lorenz-type curves for MIT harm severity ratings by pathway: (a) direct, (b) indirect, and (c) inferred. The legend is shared across panels and uses the same colour for each harm category throughout.}
\label{fig:mit_lorenz_pathways}
\end{figure}
\FloatBarrier

Figure~\ref{fig:mit_lorenz_pathways} shows the ordered Lorenz-type curves used to calculate the pathway-specific AIH scores, with panels (a), (b), and (c) corresponding to the direct, indirect, and inferred pathways, respectively. The curves have relatively few visible points because the MIT ratings take only six ordered values, from 0 to 5. Each point corresponds to one cumulative rating level, not to an individual incident. For example, in the direct pathway, financial harm has 1166 of the 1498 incidents at rating 1 and only 10 incidents at rating 5. The curve therefore moves quickly through a large share of low-rated incidents before reaching the highest severity rank. This explains why the direct financial AIH value is moderate ($AIH=0.3063$): there is some mass at higher severity levels, but most observations remain concentrated at low ratings. Malicious content, by contrast, has a larger relative shift toward higher direct severity ratings, which is why its curve produces the highest direct AIH score. The same logic applies pathway by pathway: AIH summarizes how far the ordered severity distribution moves toward higher ratings without treating the rating labels as cardinal numbers.

\subsection{Multivariate analysis: Lorenz zonoids and Gini indices}

The main analysis combines the three MIT severity pathways rather than evaluating them separately. For each harm category, we construct severity-level frequency matrices in which the rows correspond to the five positive severity levels and the columns correspond to the Direct, Indirect, and Inferred pathways. The empirical Lorenz zonoid is then computed from these frequency matrices using the subset construction described above. This step preserves the joint pathway structure of the MIT annotations and allows concentration to be evaluated as a multidimensional property of each harm category.

\begin{figure}[htbp]
    \centering
    \includegraphics[width=0.4\textwidth]{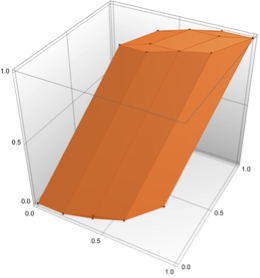}
    \hfill
    \includegraphics[width=0.4\textwidth]{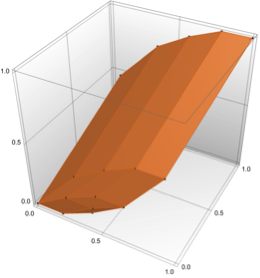}

    \vspace{0.5cm}

    \includegraphics[width=0.4\textwidth]{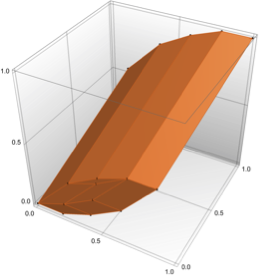}
    \caption{Empirical Lorenz zonoids for Privacy harms under the
    Direct--Indirect pathway pair (upper left), the Direct--Inferred
    pathway pair (upper right), and the Indirect--Inferred pathway pair
    (lower panel).}
    \label{ZonoidPrivacy}
\end{figure}

\begin{figure}[htbp]
    \centering
    \includegraphics[width=0.4\textwidth]{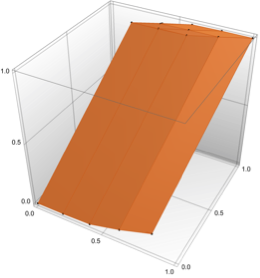}
    \hfill
    \includegraphics[width=0.4\textwidth]{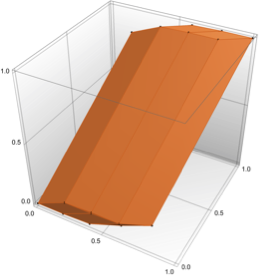}

    \vspace{0.5cm}

    \includegraphics[width=0.4\textwidth]{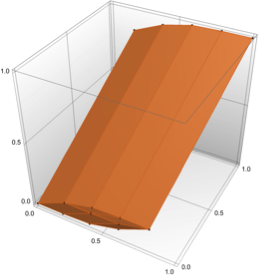}
    \caption{Empirical Lorenz zonoids for Physical harms under the
    Direct--Indirect pathway pair (upper left), the Direct--Inferred
    pathway pair (upper right), and the Indirect--Inferred pathway pair
    (lower panel).}
    \label{ZonoidPhysical}
\end{figure}

\begin{figure}[htbp]
    \centering
    \includegraphics[width=0.8\textwidth]{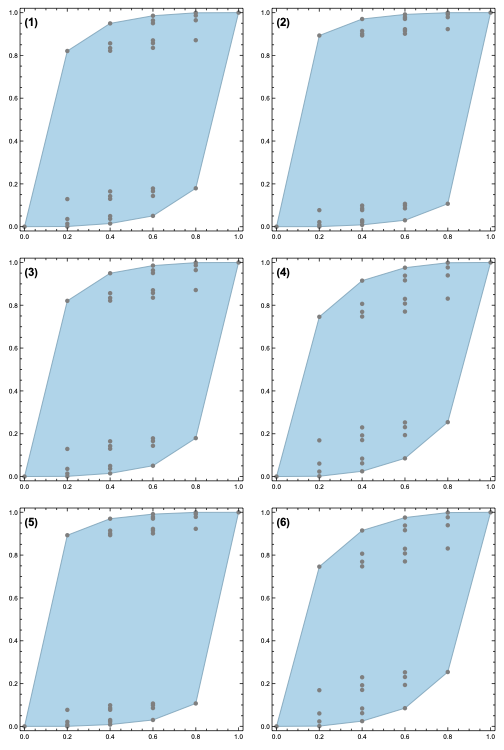}
    \caption{Pooled marginal Lorenz zonoids for all MIT harm categories.
    The six panels show: (1) the Direct margin of the Direct--Indirect
    pair; (2) the Indirect margin of the Direct--Indirect pair;
    (3) the Direct margin of the Direct--Inferred pair;
    (4) the Inferred margin of the Direct--Inferred pair;
    (5) the Indirect margin of the Indirect--Inferred pair; and
    (6) the Inferred margin of the Indirect--Inferred pair.}
    \label{UnivariateZ}
\end{figure}

\FloatBarrier

Figures~\ref{ZonoidPrivacy} and~\ref{ZonoidPhysical} present
category-specific empirical Lorenz zonoids for Privacy and Physical
harms. Each figure contains the three possible bivariate configurations
of the severity pathways: Direct--Indirect, Direct--Inferred, and
Indirect--Inferred. Because the empirical Lorenz zonoid of a
two-dimensional response is represented in $\mathbb{R}^{3}$, each
pathway pair produces a three-dimensional geometric object.

Figure~\ref{UnivariateZ} instead reports the pooled marginal zonoids
obtained from all harm categories. Its six panels correspond to the two
univariate margins associated with each of the three pathway pairs.
Figures~\ref{ZonoidPrivacy} and~\ref{ZonoidPhysical} therefore
illustrate the category-specific joint geometry of the severity
pathways, whereas Figure~\ref{UnivariateZ} summarizes their pooled
marginal structure across the MIT harm categories.

Table~\ref{tab:mit_concentration_indices} reports the multivariate
concentration indices calculated from the final MIT severity-level
frequency matrices. For each harm category, the empirical matrix
$A$ has five rows, corresponding to severity levels 1--5, and three
columns, corresponding to the Direct, Indirect, and Inferred pathways.

The relative distance-Gini $R_D(F_A)$ in
Formula~(\ref{gini3}) is used directly. For comparability, we denote
the volume-Gini index $M_V(F_A)$, normalized by the last upper bound
in Formula~(\ref{bound2}), as $\widetilde{M}_V(F_A)$. The
distance-Gini mean difference $M_D(F_A)$ in Formula~(\ref{gini1})
is also reported as an unnormalized reference measure. The two main
scale-free and directly comparable indices are therefore
$R_D(F_A)$ and $\widetilde{M}_V(F_A)$.

\begin{table}[H]
\centering
\scriptsize
\renewcommand{\arraystretch}{1.2}
\setlength{\tabcolsep}{3pt}

\begin{tabularx}{\textwidth}{
@{}
>{\raggedright\arraybackslash}p{0.22\textwidth}
>{\centering\arraybackslash}X
>{\centering\arraybackslash}X
>{\centering\arraybackslash}X
@{}
}
\toprule
\textbf{MIT harm category}
&
\shortstack[c]{
\textbf{Relative distance-Gini}\\
$\boldsymbol{R_D(F_A)}$
}
&
\shortstack[c]{
\textbf{Normalized volume-Gini}\\
$\boldsymbol{\widetilde{M}_V(F_A)}$
}
&
\shortstack[c]{
\textbf{Distance-Gini}\\
\textbf{mean difference}\\
$\boldsymbol{M_D(F_A)}$
}
\\
\midrule
Environmental
    & 0.4612 & 0.6584 & 137.6886 \\
Infrastructure
    & 0.4592 & 0.6557 & 137.0965 \\
Property
    & 0.4500 & 0.6428 & 134.4053 \\
Physical
    & 0.4377 & 0.6258 & 130.8125 \\
Democracy
    & 0.4196 & 0.6001 & 125.3766 \\
Financial
    & 0.3909 & 0.5423 & 116.8274 \\
Differential Treatment
    & 0.3870 & 0.5538 & 115.7154 \\
Privacy
    & 0.3784 & 0.5374 & 113.1207 \\
Civil Rights
    & 0.3701 & 0.5250 & 110.6962 \\
Malicious Content
    & 0.3592 & 0.4985 & 107.3908 \\
\bottomrule
\end{tabularx}

\caption{Multivariate concentration indices for the MIT harm
categories. The relative distance-Gini $R_D(F_A)$ is defined in
Formula~(\ref{gini3}). The normalized volume-Gini
$\widetilde{M}_V(F_A)$ is obtained by calculating $M_V(F_A)$ from
Formula~(\ref{volumeGini}) using the column-normalized pathway
frequency shares and dividing it by the last upper bound in
Formula~(\ref{bound2}). The distance-Gini mean difference $M_D(F_A)$
from Formula~(\ref{gini1}) is reported as an unnormalized reference
measure.}
\label{tab:mit_concentration_indices}
\end{table}
\begin{figure}[H]
\centering
\includegraphics[width=1\textwidth]{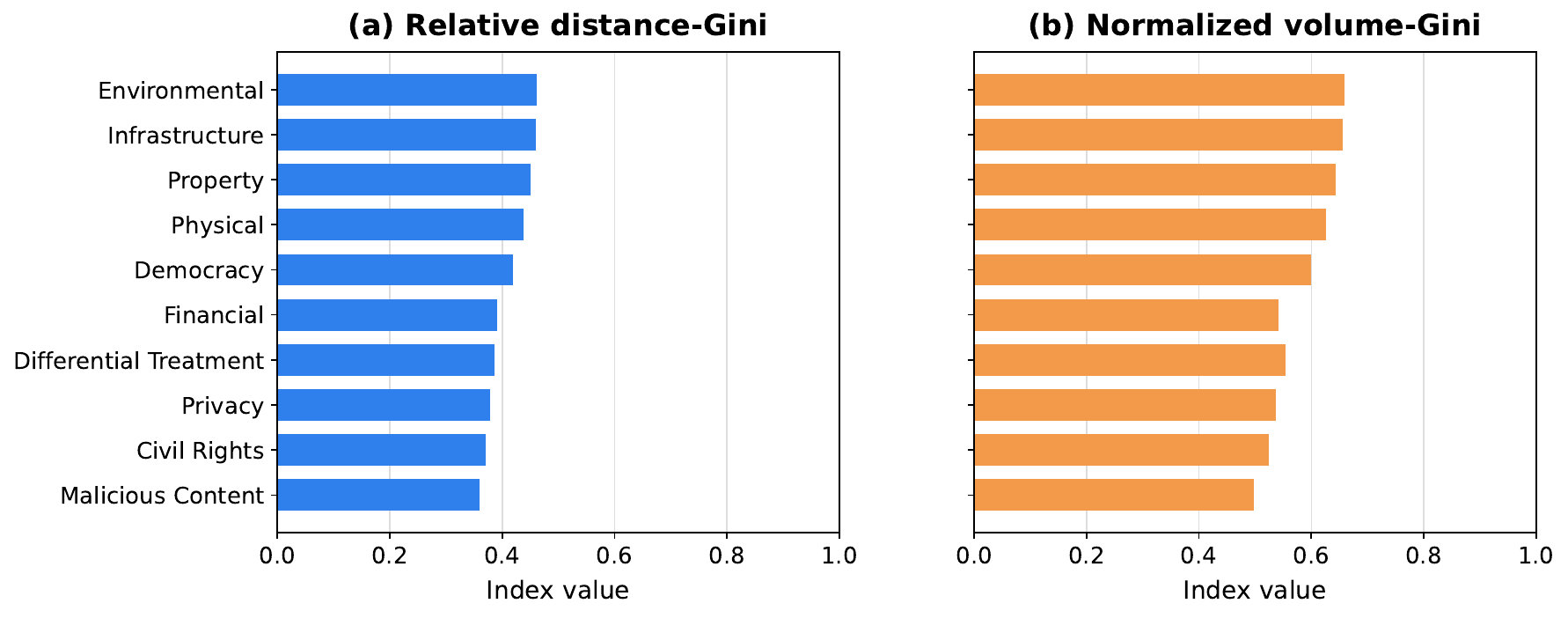}
\caption{Multivariate concentration indices by MIT harm category. The left panel reports the relative distance-Gini $R_D(F_A)$ from Formula~(\ref{gini3}), while the right panel reports the normalized volume-Gini $\widetilde{M}_V(F_A)$ obtained from Formula~(\ref{volumeGini}) and the upper bound in Formula~(\ref{bound2}).}
\label{fig:mit_concentration_indices}
\end{figure}

Table~\ref{tab:mit_concentration_indices} and Figure~\ref{fig:mit_concentration_indices} show that the two scale-free concentration indices produce broadly similar
rankings of the harm categories. Environmental and infrastructure
harms obtain the highest values under both $R_D(F_A)$ and
$\widetilde{M}_V(F_A)$, followed by property, physical, and
democracy-related harms. These categories therefore exhibit the
strongest concentration in their joint severity-level frequency
distributions across the Direct, Indirect, and Inferred pathways. Accordingly, under the multidimensional concentration criterion adopted
in this study, environmental, infrastructure, property, physical, and
democracy-related harms emerge as the most prominent categories because
they attain the highest values under both $R_D(F_A)$ and
$\widetilde{M}_V(F_A)$. In this specific sense, the indices can support
the identification of harm categories that may require greater
attention in mitigation planning. This interpretation is specific to multidimensional concentration.
Higher values of $R_D(F_A)$ and $\widetilde{M}_V(F_A)$ do not by
themselves indicate greater absolute severity, higher prevalence, a
larger number of affected individuals, or greater overall societal
importance. Mitigation priorities should therefore also consider the
pathway-specific ordinal AIH values and substantive information about
the incidents and affected stakeholders.

Malicious content, civil rights, and privacy obtain values near the
lower end of both rankings. Financial and differential-treatment harms
occupy intermediate positions and exchange order across the two
indices. The unnormalized distance-Gini mean difference $M_D(F_A)$
produces an ordering that is generally consistent with the two
scale-free measures, although its magnitude depends on the scale of the
underlying frequency matrices and should therefore be interpreted only
as a reference value.

Some categories with relatively high pathway-specific AIH values in
Table~\ref{tab:mit_pathway_aih} receive lower multivariate
concentration-index values in Table~\ref{tab:mit_concentration_indices}.
This difference is expected because the pathway-specific AIH analysis
evaluates the ordered severity distribution of each pathway
independently. By contrast, the multivariate concentration indices
summarize the joint Direct--Indirect--Inferred frequency structure of
each harm category.

Although Malicious Content has the highest direct AIH value, it has one of the lowest multivariate concentration-index values. This is because AIH evaluates each pathway separately as an ordered severity distribution, whereas the zonoid indices evaluate the joint Direct--Indirect--Inferred severity-frequency matrix for each harm category.

\section{Conclusions}
In this paper, we have shown how multidimensional Lorenz zonoids and
the related multivariate Gini indices can be calculated empirically.
The methodology was applied to ten AI-harm categories whose severity
is recorded ordinally across the Direct, Indirect, and Inferred
pathways in the MIT AI Incident Tracker.

The results show that environmental, infrastructure, property, physical, and democracy-related harms attain the highest values under the relative distance-Gini $R_D(F_A)$ and the normalized volume-Gini $\widetilde{M}_V(F_A)$. Under the multidimensional concentration criterion adopted in this study, these categories therefore exhibit the strongest concentration in their joint severity-frequency structures and may warrant closer examination when mitigation priorities are considered.

The multidimensional concentration indices complement the pathway-specific AIH analysis. AIH captures how ordinal severity is distributed within each pathway, whereas the multivariate indices characterize the joint frequency structure across the three pathways. This distinction explains why categories with high pathway-specific AIH values do not necessarily obtain high
multivariate concentration values.

Multidimensional Lorenz zonoids and Gini indices provide a transparent way to examine the joint structure of multidimensional AI harm annotations. Used together with ordinal severity measures, prevalence information and substantive stakeholder analysis, they can contribute to more informed AI-harm assessment and mitigation planning.

\section*{Acknowledgements}

The authors thank Peter Slattery and Simon Mylius from the Massachusetts Institute of Technology, as the paper relies on two data sources: the incident reports, collected and curated by the AI Incident Database; and the analysis, risk classification, and severity assessments from the MIT AI Incident Tracker.

J.M. Sarabia acknowledges the financial support of the R\&D\&I project Ref. PID2024-156871NBI00, funded by MICIU/AEI/10.13039/501100011033/ FEDER,EU.

\newpage

\bibliographystyle{Chicago.bst}
\bibliography{Bibliography}

@article{koshevoy1996lorenz,
  title={The Lorenz zonoid of a multivariate distribution},
  author={Koshevoy, Gleb and Mosler, Karl},
  journal={Journal of the American Statistical Association},
  volume={91},
  number={434},
  pages={873--882},
  year={1996},
  publisher={Taylor \& Francis}
}

@article{koshevoy1997zonoid,
  title={Zonoid trimming for multivariate distributions},
  author={Koshevoy, Gleb and Mosler, Karl},
  journal={The Annals of Statistics},
  volume={25},
  number={5},
  pages={1998--2017},
  year={1997},
  publisher={Institute of Mathematical Statistics}
}

@book{arnold2018majorization,
  title={Majorization and the Lorenz order with applications in applied mathematics and economics},
  author={Arnold, Barry C and Sarabia, Jos{\'e} M},
  volume={7},
  year={2018},
  publisher={Springer}
}

@article{koshevoy1997multivariate,
  title={Multivariate gini indices},
  author={Koshevoy, Gleb A and Mosler, Karl},
  journal={Journal of Multivariate Analysis},
  volume={60},
  number={2},
  pages={252--276},
  year={1997},
  publisher={Elsevier}
}

@article{Giudici2023safe,
  title={SAFE Artificial Intelligence in finance},
  author={Giudici, Paolo and Raffinetti, Emanuela},
  journal={Finance Research Letters},
  volume={56},
  pages={104088},
  year={2023},
  publisher={Elsevier}
}

@article{Giudici2024artificial,
  title={Artificial Intelligence risk measurement},
  author={Giudici, Paolo and Centurelli, Mattia and Turchetta, Stefano},
  journal={Expert Systems with Applications},
  volume={235},
  pages={121220},
  year={2024},
  publisher={Elsevier}
}

@article{Liu2022trustworthy,
  title={Trustworthy ai: A computational perspective},
  author={Liu, Haochen and Wang, Yiqi and Fan, Wenqi and Liu, Xiaorui and Li, Yaxin and Jain, Shaili and Liu, Yunhao and Jain, Anil and Tang, Jiliang},
  journal={ACM Transactions on Intelligent Systems and Technology},
  volume={14},
  number={1},
  pages={1--59},
  year={2022},
  publisher={ACM New York, NY}
}

@article{Kaur2022trustworthy,
  title={Trustworthy artificial intelligence: a review},
  author={Kaur, Davinder and Uslu, Suleyman and Rittichier, Kaley J and Durresi, Arjan},
  journal={ACM computing surveys (CSUR)},
  volume={55},
  number={2},
  pages={1--38},
  year={2022},
  publisher={ACM New York, NY}
}

@article{Cheng2021socially,
  title={Socially responsible ai algorithms: Issues, purposes, and challenges},
  author={Cheng, Lu and Varshney, Kush R and Liu, Huan},
  journal={Journal of Artificial Intelligence Research},
  volume={71},
  pages={1137--1181},
  year={2021}
}

@article{smuha2025european,
  author  = {Smuha, Nathalie A. and Yeung, Karen},
  title   = {The {European Union}'s {AI} {Act}: beyond motherhood and apple pie?},
  year    = {2025},
    publisher={{The Cambridge Handbook of the Law, Ethics and Policy of Artificial Intelligence. Cambridge Law Handbooks.}},
    pages = {228-258},
  isbn = {9781009367813},
doi = {10.1017/9781009367783.015}
}

@article{giudici2021cyber,
  author  = {Giudici, Paolo and Raffinetti, Emanuela},
  title   = {Explainable {AI} methods in cyber risk measurement},
  journal = {Quality and Reliability Engineering International},
  year    = {2023}
}

@article{novelli2024taking,
  author  = {Novelli, Claudio and Casolari, Federico and Rotolo, Antonino and Taddeo, Mariarosaria and Floridi, Luciano},
  title   = {Taking {AI} risks seriously: a new assessment model for the {AI} {Act}},
  journal = {{AI} \& Society},
  volume  = {39},
  number  = {5},
  pages   = {2493--2497},
  year    = {2024},
  doi     = {10.1007/s00146-023-01723-z}
}

@article{hoffman2023outside,
  author  = {Hoffman, Fred and Kesharwani, Rajkamal and Maynard, Jacob},
  title   = {How an outside perspective can help an organization enhance their supply chain risk mitigation strategy},
  journal = {Issues in Information Systems},
  volume  = {24},
  number  = {1},
  year    = {2023}
}

@article{lancaster2024s,
  author  = {Lancaster, Caitlin M and Schulenberg, Kelsea and Flathmann, Christopher and McNeese, Nathan J and Freeman, Guo},
  title   = {``{It}'s Everybody's Role to Speak Up... But Not Everyone Will'': Understanding {AI} Professionals' Perceptions of Accountability for {AI} Bias Mitigation},
  journal = {ACM Journal on Responsible Computing},
  volume  = {1},
  number  = {1},
  pages   = {1--30},
  year    = {2024},
  doi     = {10.1145/3630106}
}

@article{jariwala2024comparative,
  author  = {Jariwala, Mayur},
  title   = {A Comparative Analysis of the {EU} {AI} {Act} and the {Colorado} {AI} {Act}: Regulatory Approaches to {Artificial Intelligence} Governance},
  journal = {International Journal of Computer Applications},
  volume  = {186},
  pages   = {23--29},
  year    = {2024},
  doi     = {10.5120/ijca2024923954}
}

@incollection{robinson2024likert,
  author    = {Robinson, John},
  title     = {Likert scale},
  booktitle = {Encyclopedia of Quality of Life and Well-Being Research},
  pages     = {3917--3918},
  isbn      = {978-3-031-17299-1},
  doi       = {10.1007/978-3-031-17299-1_1654},
  year      = {2024},
  publisher = {Springer}
}

@article{VEI2026104587,
title = {AI Harmonics: A human-centric and harms severity-adaptive AI risk assessment framework},
journal = {Artificial Intelligence},
volume = {358},
pages = {104587},
year = {2026},
issn = {0004-3702},
doi = {https://doi.org/10.1016/j.artint.2026.104587},
url = {https://www.sciencedirect.com/science/article/pii/S000437022600113X},
author = {Sofia Vei and Paolo Giudici and Pavlos Sermpezis and Athena Vakali and Adelaide Emma Bernardelli}
}

@article{grothe2022multivariate,
  title={A multivariate extension of the Lorenz curve based on copulas and a related multivariate Gini coefficient},
  author={Grothe, Oliver and K{\"a}chele, Fabian and Schmid, Friedrich},
  journal={The Journal of Economic Inequality},
  volume={20},
  number={3},
  pages={727--748},
  year={2022},
  publisher={Springer}
}

@book{marshall1979inequalities,
  author={Marshall, A. W. and Olkin, I.  and Arnold, B.C.},
    title = {Inequalities: theory of majorization and its applications (second edition)},
   publisher={Springer},
     year={2011}
     }

@article{arnold2011inequality,
  title={Inequality and majorization: Robin Hood in unexpected places},
  author={Arnold, Barry C},
  journal={Calcutta Statistical Association Bulletin},
  volume={63},
  number={1-4},
  pages={71--80},
  year={2011},
  publisher={SAGE Publications Sage India: New Delhi, India}
}

@misc{arnold1983pareto,
  title={Pareto Distributions, International Cooperative Publishing House},
  author={Arnold, BC},
  year={1983},
  publisher={Fairland Maryland}
}

@article{taguchi1972two,
  title={On the two-dimensional concentration surface and extensions of concentration coefficient and pareto distribution to the two dimensional case—I: On an application of differential geometric methods to statistical analysis},
  author={Taguchi, Tokio},
  journal={Annals of the Institute of Statistical Mathematics},
  volume={24},
  number={1},
  pages={355--381},
  year={1972},
  publisher={Springer}
}

@article{lunetta1972di,
  title={Di un indice di cocentrazione per variabili statistische doppie},
  author={Lunetta, G},
  journal={Annali della Facolt{\'a} di Economia e Commercio dell Universit{\'a} di Catania, A},
  volume={18},
  year={1972}
}

@article{auricchio2026rank,
  title={On rank graduation metrics for high--dimensional ordinal data},
  author={Auricchio, Gennaro and Bernardelli, Adelaide Emma and Giudici, Paolo and Toscani, Giuseppe},
  journal={Mathematical Models and Methods in Applied Sciences},
  volume={36},
  number={08},
  pages={1679--1713},
  year={2026},
  publisher={World Scientific}
}

@article{koshevoy1995multivariate,
  title={Multivariate lorenz majorization},
  author={Koshevoy, Gleb},
  journal={Social Choice and Welfare},
  pages={93--102},
  year={1995},
  publisher={JSTOR}
}

@article{fan2024multidimensional,
  title={Multidimensional inequality measurement via optimal transport},
  author={Fan, Yanqin and Henry, Marc and Pass, Brendan and Rivero, Jorge A},
  journal={Review of Economics and Statistics},
  pages={1--45},
  year={2024},
  publisher={MIT Press 255 Main Street, 9th Floor, Cambridge, Massachusetts 02142, USA~…}
}

@article{hallin2025multiple,
  title={Multiple-attribute Lorenz functions and Gini indices: A measure transportation approach},
  author={Hallin, Marc and Mordant, Gilles},
  journal={Journal of Business \& Economic Statistics},
  volume={43},
  number={4},
  pages={1092--1104},
  year={2025},
  publisher={Taylor \& Francis}
}

\end{document}